\documentclass[10pt]{article}

\usepackage{doublespace}
\usepackage{note}
\usepackage{epsfig}
\usepackage{latexsym}

\begin{document}
\title{Oblivious remote state preparation}
\author{Debbie W. Leung$^1$ and Peter W. Shor$^2$} 
\address{
$^1$IBM T.J.~Watson Research Center, P.O.~Box 218, 
Yorktown Heights, NY 10598, USA \\
$^2$AT\&T Labs -- Research, Florham Park, NJ 07932, USA \\[1.5ex]
}
\maketitle
\centerline{January 03, 2001}

\vspace*{-4ex} 

\begin{abstract}

We consider remote state preparation protocols for a set of pure
states whose projectors form a basis for operators acting on
the input Hilbert space.
If a protocol 
(1) uses only forward communication and entanglement,
(2) deterministically prepares an exact copy of the state, and 
(3) does so {\em obliviously}---without leaking further information 
    about the state to the receiver---
then the protocol can be modified to require from the sender {\em only} 
a single specimen of the state.   
Furthermore, the original protocol and the modified protocol use the 
same amount of classical communication.  
Thus, under the three conditions stated, remote state preparation
requires at least as much classical communication as teleportation, as
Lo has conjectured [PRA 62 (2000) 012313], which is twice the expected
classical communication cost of some existing nonoblivious protocols.

\end{abstract}

\vspace*{4ex} 

Teleportation~\cite{Bennett93} is a protocol that enables a quantum
state to be transmitted from a sender (``Alice'') to a receiver
(``Bob'') using only quantum entanglement and classical communication.
To communicate any state in a $2$-dimensional Hilbert space (qubit),
it suffices for Alice and Bob to share $1$ EPR state (ebit), 
${1 \over \sqrt{2}} (|00\>+|11\>)$ and for Alice to send $2$ classical 
bits (cbits) to Bob.  
These resources are also necessary, because teleportation preserves
the entanglement shared between the transmitted state and any other
systems, and can be used to share entanglement or to perform
superdense coding~\cite{Bennett92}.  
The procedure for teleportation does not depend on the transmitted
state, with a trivial consequence that pure states cannot be sent with
fewer resources.

More recently, Lo~\cite{Lo00rsp} studied methods to transmit quantum
states using entanglement and classical communication when the sender
has knowledge of the transmitted state.  This communication task is
called remote state preparation (RSP).  
RSP protocols more economical than teleportation were found for certain 
ensembles of pure states.
\footnote{An ensemble is the set of quantum messages, endowed with a
probability distribution, that the sender may send.}
The suggested possibility to trade off the two resources were studied
in detail~\cite{Bennett00,Devetak01}.

Resource lower bounds for RSP of pure states are generally
difficult to establish.  
Unlike teleportation, RSP of pure states need not preserve the
entanglement of the transmitted system with other systems, so that
neither lower bounds for teleportation applies.
For instance, the classical communication cost for RSP of an arbitrary
pure $d$-dimensional state is only lower bounded by $\log d$ cbits
(Holevo's bound~\cite{Holevo73}), in contrast to the $2 \log d$ cbits
required for teleportation.
In Ref.~\cite{Lo00rsp}, Lo conjectured that $2 \log d$ cbits are
indeed necessary for RSP.
But Ref.~\cite{Bennett00} found probabilistic RSP protocols with an
expected classical communication cost saturating Holevo's bound. 
However Ref.~\cite{Bennett00} suggested that Lo's conjecture may still
hold in certain circumstances, such as when the protocol uses a
constant (non-probabilistic) amount of classical communication, or
when it leaks no extra information to Bob about the state being
prepared, beyond that already contained in the state itself.

In this paper, we prove a stronger result that implies Lo's conjecture
under circumstances to be defined.
%
%
%
We use the term ``generic ensemble'' to describe an ensemble of states
whose density matrices form a basis of operators acting on some (input)  
Hilbert space.
We say that an RSP protocol is oblivious to Bob if he obtains no more
information about the prepared state than is contained in the single
specimen, even if he deviates from the protocol.
A protocol is called faithful if it is exact and deterministic.
Finally, a protocol is said to be oblivious to Alice if it requires
from her only a specimen of the transmitted state, but not her
knowledge of it.
With these definitions, we can state our result: 
\begin{quote}
If an RSP protocol for a general ensemble of pure states uses only
forward communication and entanglement, and is faithful and oblivious
to Bob, then it can be modified to be oblivious to Alice at no extra
classical communication cost. 
\end{quote}
An immediate corollary is that such an RSP protocol uses at least as
much classical communication as required in teleportation.
Our work also elicits conditions under which RSP is suboptimal to
teleportation, and provides insights on how knowledge of the prepared
states enables resource tradeoff in RSP.

Our result follows from an explicit procedure to convert a faithful
RSP protocol oblivious to Bob and using no back communication to a
protocol oblivious to Alice.
Suppose the original RSP protocol transmits any state $\phi$ drawn from
a generic ensemble of $d$-dimensional states.
The most general faithful protocol without back communication is 
represented by the circuit:

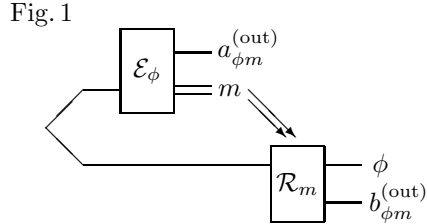
\begin{figure}[ht]
\setlength{\unitlength}{0.5mm}
\centering
\begin{picture}(120,60)
\put(0,50){\makebox(18,10){Fig.\,1}}
\put(20,15){\line(1,0){50}}
\put(70,0){\framebox(14,20){${\cal R}_m$}}
\put(84,5){\line(1,0){10}}
\put(84,15){\line(1,0){10}}
\put(95,0){\makebox(18,10){$b^{({\rm out})}_{\phi m}$}}
\put(94,10){\makebox(10,10){$\phi$}}
\put(10,25){\line(1,1){10}}
\put(10,25){\line(1,-1){10}}
\put(20,35){\line(1,0){10}}
\put(30,29){\framebox(14,22){${\cal E}_\phi$}}
\put(44,45){\line(1,0){10}}
\put(44,36){\line(1,0){10}}
\put(44,34){\line(1,0){10}}
\put(54,30){\makebox(10,10){$m$}}
\put(55,41){\makebox(18,10){$a^{({\rm out})}_{\phi m}$}}
\put(64,35){\vector(1,-1){12.5}}
\put(64,32.5){\vector(1,-1){10}}
\end{picture}
\label{fig:rsp1}
\end{figure}

In the above diagram, the entangled state shared by Alice and Bob 
is a maximally entangled state in $2$ $d'$-dimensional systems, 
$|\Phi_{d'}\> = {1 \over \sqrt{d'}} 
\lpm |11\> + \cdots + |d'\! d' \, \> \rpm$.  We do not require $d = d'$.
The most general action of Alice is to apply to her half of
$|\Phi_{d'}\>$ a trace preserving quantum operation ${\cal E}_\phi$,
parameterized by the input $\phi$ to reflect possible use of her
knowledge of it.
\footnote{In principle, Alice can input an ancilla $a_\phi^{({\rm
in})}$ to ${\cal E}_\phi$, but the effect is just a redefinition of
${\cal E}_\phi$.}
Since the communication is classical, ${\cal E}_\phi$ should output
some classical message $m$ to be sent to Bob, with probability
$(p_\phi)_m$.  Note that $\sum_m (p_\phi)_m = 1$.  The remaining
classical or quantum output is collectively represented by $a_{\phi
m}^{\rm (out)}$.
Since the protocol is faithful, there exists a ``recovery'' 
procedure for Bob that depends on $m$ but not on $\phi$. 
The most general procedure is a {\em trace preserving} quantum
operation ${\cal R}_m$ acting on his half of $|\Phi_{d'}\>$.   
\footnote{Again, his possible use of an ancilla $b_m^{\rm (in)}$ 
can be replaced by redefining ${\cal R}_m$.} 
This procedure always outputs a copy of $\phi$, and some extra
output $b_{\phi m}^{\rm (out)}$.

We now simplify the above circuit.  
Since the prepared state $\phi$ is pure, it is unaffected if
$a_\phi^{(out)}$ is traced out.  Thus Alice's operation ${\cal
E}_\phi$, with only classical output, is just a POVM.
On the other hand, since Bob's operation ${\cal R}_m$ is trace 
preserving, it can be implemented by attaching a pure ancilla $|0\>$ 
and applying a joint unitary operation $U_m$.  
The simplified circuit is given by: 

\clearpage
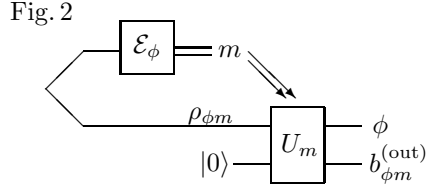
\begin{figure}[ht]
\setlength{\unitlength}{0.5mm}
\centering
\begin{picture}(120,50)
\put(0,40){\makebox(18,10){Fig.\,2}}
\put(20,15){\line(1,0){50}}
\put(70,0){\framebox(14,20){$\,U_m$}}
\put(84,5){\line(1,0){10}}
\put(84,15){\line(1,0){10}}
\put(95,0){\makebox(18,10){$b^{({\rm out})}_{\phi m}$}}
\put(94,10){\makebox(10,10){$\phi$}}
\put(48,12.5){\makebox(12,10){$\rho_{\phi m}$}}
\put(60,5){\line(1,0){10}}
\put(50,0){\makebox(10,10){$|0\rangle$}}
\put(10,25){\line(1,1){10}}
\put(10,25){\line(1,-1){10}}
\put(20,35){\line(1,0){10}}
\put(30,29){\framebox(14,14){${\cal E}_\phi$}}
\put(44,36){\line(1,0){10}}
\put(44,34){\line(1,0){10}}
\put(54,30){\makebox(10,10){$m$}}
\put(64,35){\vector(1,-1){12.5}}
\put(64,32.5){\vector(1,-1){10}}
\end{picture}
\label{fig:rsp2}
\end{figure}

where $\rho_{\phi m}$ denotes the state of Bob's half of the shared
system given the message $m$.

We now apply the oblivious condition. 
This implies $b^{\rm (out)}_{\phi m}$ and $(p_\phi)_m$ are independent
of $\phi$ and can be written as $b^{\rm (out)}_{m}$ and $p_m$
respectively; otherwise Bob can gain information about the identity of
$\phi$ without disturbing his single specimen of it, violating the
no-imprinting condition~\cite{Bennett94} for a generic ensemble of
states.  We obtain an expression for $\rho_{\phi m}$ using 
the state change due to $U_m$, 
\be 
 	\rho_{\phi m} = {\rm Tr}_{\it 2} \, 
	\lbm U_m^\dag \lpm \phi \ot b_m^{\rm (out)} \rpm U_m \rbm 
\label{eq:rhophim}
\,, 
\ee
where ${\rm Tr}_{\it 2}$ denotes the tracing of the second tensor 
component. 
Throughout the paper, an operation acts on the subsystem labelled by
an italicized numerical subscript; the ordering is based on the equation 
containing the operation.  

We obtain another identity by describing in two different ways
the state in Bob's half of the shared system before receiving the
message $m$, 
\be
	 \sum_m p_m \rho_{\phi m} = {I \over d'} \,.  
\label{eq:before} 
\ee
Substituting \eq{rhophim} into \eq{before},  
\be
	\sum_{m} p_m \lbm {\rm Tr}_{\it 2} \, 
	U_m^\dag \lpm \phi \ot b_m^{(out)} \rpm U_m \rbm 
	= {I \over d'} 
\label{eq:norm}
\,.  
\ee
Equating the left side of \eq{norm} to ${\cal F}(\phi)$ defines an
operation ${\cal F}$ acting on $\phi$.  Since the set of all possible
$\phi$ forms a generic ensemble, ${\cal F}$ is defined for all
operators acting on the input Hilbert space, and it is simply the
randomizing operation, ${\cal F}(\cdot) = {I \over d'}$.

We are now ready to describe a modified protocol in which Bob receives
the same classical and quantum outputs as in the original RSP protocol
but Alice applies a $\phi$ independent measurement $\cal M$ to a single
specimen of $\phi$ and her half of $|\Phi_{d'}\>$:
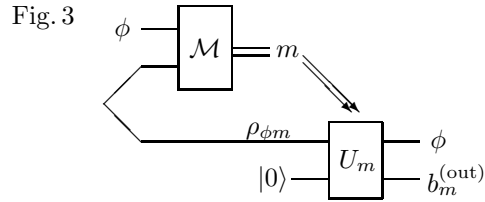
\begin{figure}[ht]
\setlength{\unitlength}{0.5mm}
\centering
\begin{picture}(120,53)
\put(-15,43){\makebox(18,10){Fig.\,3}}
\put(20,15){\line(1,0){50}}
\put(70,0){\framebox(14,20){$\,U_m$}}
\put(84,5){\line(1,0){10}}
\put(84,15){\line(1,0){10}}
\put(95,0){\makebox(18,10){$b^{({\rm out})}_m$}}
\put(94,10){\makebox(10,10){$\phi$}}
\put(48,12.5){\makebox(12,10){$\rho_{\phi m}$}}
\put(60,5){\line(1,0){10}}
\put(50,0){\makebox(10,10){$|0\rangle$}}
\put(10,40){\makebox(10,10){$\phi$}}
\put(20,45){\line(1,0){10}}
\put(10,25){\line(1,1){10}}
\put(10,25){\line(1,-1){10}}
\put(20,35){\line(1,0){10}}
\put(30,29){\framebox(14,22){${\cal M}$}}
\put(44,40){\line(1,0){10}}
\put(44,38){\line(1,0){10}}
\put(54,34){\makebox(10,10){$m$}}
\put(63,38){\vector(1,-1){15}}
\put(63,36){\vector(1,-1){13}}
\end{picture}
\label{fig:rsp3}
\end{figure}

The POVM elements of ${\cal M}$ are given by
\be
	M_m = d d' \, p_m {\rm Tr}_{\it 3} 
	\lpm I_{\it 1} \ot U_{m {\it 23}}^T \rpm 
	\lpm |\Phi_d\>\<\Phi_d|_{\it 12} 
	\ot b_{m {\it 3}}^{{\rm (out)}\,T} \rpm 
	\lpm I_{\it 1} \ot U_{m {\it 23}}^* \rpm
\label{eq:mm}
\ee
We first verify that $\{M_m\}$ is a POVM acting on a $d \times d'$ system.
According to \eq{rhophim}, $I_{\it 1} \ot U_{m {\it 23}}^T$ maps an
operator acting on a $d \times d \times {\rm dim}(b^{\rm (out)}_{m})$
system to one acting on a $d \times d' \times 1$ system.  The ${\rm
Tr}_{\it 3}$ in \eq{mm} then ensures $M_m$ acts on a $d \times d'$
system.
Each $M_m$ is manifestly positive.
Furthermore, let ${\cal I}$ denote the identity operation.  Using \eq{norm}, 
\be
      \sum_m M_m^T = d d' ({\cal F} \ot {\cal I})    
	(|\Phi_d\>\<\Phi_d|) = I \ot I
\,,
\ee
so that $\{M_m\}$ is indeed a POVM.
Note that ${\cal M}$ is independent of $\phi$, and the modified
protocol in Fig.\,3 is indeed oblivious to Alice.
It remains to verify that the modified protocol is the same as the
original one from Bob's point of view.
Let ${\bf b}$ be the state in Bob's half of the shared system when the
measurement $\cal M$ outputs $m$, normalized by the probability of
outcome $m$.
The modified protocol creates the correct state with the correct
probability if ${\bf b} = p_m \rho_{\phi m}$.
Evaluating ${\bf b}$, 
\bea
	{\bf b} & = & {\rm Tr}_{\it 12} \lbL 
	\lpm \phi_{\it 1} \ot |\Phi_{d'}\>\<\Phi_{d'}|_{\it 23} \rpm 
	\lpm M_{m {\it 12}} \ot I_{\it 3} \rpm \rbL
\\
	& = &  \sum_{j_1\!,\,j_2} {\rm Tr}_{\it 12} \lbL {1 \over d'}  
	\lpm \phi_{\it 1} \ot |j_1\>\<j_2|_{\it 2} \ot |j_1\>\<j_2|_{\it 3} 
	\rpm  \;
	\lpm M_{m {\it 12}} \ot I_{\it 3} \rpm \rbL   
\\
	& = & \sum_{j_1\!,\,j_2} {\rm Tr} \lbL {1 \over d'} 
	\lpm \phi \ot |j_1\>\<j_2| \rpm \; 
	M_{m} \rbL |j_1\>\<j_2|
\label{eq:b0}
\,.
\eea
So, the $(j_1,j_2)$ entry of ${\bf b}$ is given by: 
\bea 
	&  & {\rm Tr} \lbL {1 \over d'} \lpm \phi^T \ot |j_2\>\<j_1| \rpm 
				\; M_{m}^T \rbL
\label{eq:b1} 
\\
	& = & d \, p_m \, 
	{\rm Tr} \lbL \lpm \phi^T \ot |j_2\>\<j_1| \ot I \rpm \;  
	\lpm I_{\it 1} \ot U_{m {\it 23}}^\dag \rpm \; 
	\lpm |\Phi_d\>\<\Phi_d|_{\it 12} \ot b_{m {\it 3}}^{\rm (out)} \rpm \;
	\lpm I_{\it 1} \ot U_{m {\it 23}} \rpm \rbL
\label{eq:b2} 
\\
	& = & p_m \,   
	{\rm Tr} \lbL \lpm |j_2\>\<j_1| \ot I \rpm \; \;
	U_{m}^\dag \lpm \phi \ot b_{m}^{\rm (out)} \rpm U_{m} \rbL
\label{eq:b3} 
\\
	& = & p_m \,   
	{\rm Tr} \lbL |j_2\>\<j_1| \;
	{\rm Tr}_{\it 2} \lbm U_{m}^\dag  
	\lpm \phi \ot b_{m}^{\rm (out)} \rpm 
	U_{m} \rbm \rbL
\label{eq:b4} 
\\
	& = & p_m \, \lbL   
	{\rm Tr}_{\it 2} \lbm U_{m}^\dag  
	\lpm \phi \ot b_{m}^{\rm (out)} \rpm U_{m} \rbm \rbL_{j_1\!,\,j_2} 
	= ~ p_m \, [ \, \rho_{\phi m} \, ]_{j_1\!,\,j_2}
\label{eq:b5} 
\eea
Equation (\ref{eq:b1}) is obtained from \eq{b0} using ${\rm Tr} AB =
{\rm Tr} A^T B^T$.  Equation (\ref{eq:b2}) follows from substituting
\eq{mm} and from the identity ${\rm Tr} \lpm \! A \ ({\rm Tr}_{\it 2}
B_{\it 12}) \rpm = {\rm Tr} \lpm (A \ot I) \ B \rpm$.  We trace out
the first register and used $d \, {\rm Tr}_{\it 1} \lbm \phi^T_{\it 1} \,
|\Phi_d\>\<\Phi_d|_{\it 12} \rbm = \phi$ to obtain \eq{b3}.  Equation
(\ref{eq:b4}) is again due to the identity ${\rm Tr} \lpm \! A \ ({\rm
Tr}_{\it 2} B_{\it 12}) \rpm = {\rm Tr} \lpm (A \ot I) \ B \rpm$.
%
Thus ${\bf b} = p_m \rho_{\phi m}$, completing the proof of our major claim. 

We make some important observations.  
First, the modification leaves $p_m$ and therefore the classical
communication cost unchanged.  
Second, while the original RSP protocol needs not preserve the
entanglement shared between the transmitted system and other systems,
the modified protocol does.
Therefore, the modified protocol can be used for superdense coding,
implying that it requires at least as much classical communication as
teleportation.
We emphasize that we have never removed the premise that the original
protocol works only for pure states; it is the modification that makes
the modified protocol more versatile.
Putting these two observations together, the original RSP protocol
must require at least $2 \log d$ cbits, proving Lo's conjecture under
the conditions imposed on the RSP protocol.

An RSP protocol that deterministically prepares $\phi$ may use a
probabilistic amount of resources.
For example, if an RSP protocol sometimes fails to prepare $\phi$, one
can still teleport $\phi$ when the protocol fails and obtain a
deterministic protocol with probabilistic resources.
We remark that our result is applicable even when the required
resources are probabilistic.  In the general description of an RSP
protocol in Fig.\,1, the classical message $m$ may have variable
length and the extra outputs $a_{\phi m}^{\rm (out)}$ and $b_{\phi
m}^{\rm (out)}$ may contain unused entanglement.
The original and modified protocol require the same amount of
classical communication in all measures, including the worst case and
the average costs.

We believe that the current result can be extended to provide a lower
bound on the entanglement required by the original RSP protocol.
When calculating the actual entanglement consumed by the original RSP
protocol, one needs to take into account unused or 1-way distillable 
entanglement between $a_{\phi m}^{\rm (out)}$ and $b_{\phi m}^{\rm
(out)}$ in Fig.\,1 (this is especially important for the average or
amortized cost).
The simplifications leading to Fig.\,2 come at a price, since we might
have discarded recoverable entanglement, and used more entanglement in
the modified protocol.
One can avoid such problem by including some 1-way entanglement
recovery procedure as part of RSP and replacing Fig.\,2 by
\begin{figure}[ht]
\setlength{\unitlength}{0.5mm}
\centering
\begin{picture}(120,65)
\put(0,55){\makebox(18,10){Fig.\,4}}
\put(20,25){\line(1,0){50}}
\put(70,0){\framebox(15,30){${\cal U}_{m,l}$}}
\put(85,15){\line(1,0){10}}
\put(85,25){\line(1,0){10}}
\put(85,5){\line(1,0){10}}
\put(95,5){\circle*{2}}
\put(95.5,0){\makebox(10,10){$b'$}}
\put(96,10){\makebox(18,10){$b^{({\rm out})}_{m,l}$}}
\put(95,20){\makebox(10,10){$\phi$}}
\put(10,35){\line(1,1){10}}
\put(10,35){\line(1,-1){10}}
\put(20,45){\line(1,0){10}}
\put(30,39){\framebox(14,22){${\cal E}_\phi$}}
\put(44,55){\line(1,0){51}}
\put(95,55){\circle*{2}}
\put(95.5,50){\makebox(10,10){$a'$}}
\put(44,46){\line(1,0){10}}
\put(44,44){\line(1,0){10}}
\put(53,40){\makebox(16,10){$m,l$}}
\put(62,41){\vector(1,-1){10}}
\put(64,41){\vector(1,-1){10}}
\end{picture}
\end{figure}
In Fig.\,4, the recovery procedure is represented by an extra
classical message $l$ and a variable amount of recovered entanglement
in the outputs $a'$ and $b'$.
We believe that the above protocol can be made oblivious to Alice
without affecting the classical and quantum outputs, and the method
will be reported in the future.

We can extend the current result to an RSP protocol that is not
faithful.  Instead the protocol sometimes prepares an exact copy of
$\phi$ but fails with some probability $p_f$.  Alice knows when it
fails, and after Bob is informed of the failure, his half of the
shared system is left in a state $\rho_f$ independent of $\phi$.  Our
previous arguments for a faithful protocol hold almost exactly, except
now $\sum_m p_m = 1-p_f$ and Figs.\,1-3 only occur with probability
$1-p_f$.  Equations (\ref{eq:before}) and (\ref{eq:norm}) are
respectively replaced by $\sum_m p_m \rho_{\phi m} + p_f \rho_f = {I
\over d'}$ and $p_f \rho_f + \sum_{m} p_m \lbm {\rm Tr}_{\it 2} \,
U_m^\dag \lpm \phi \ot b_m^{(out)} \rpm U_m \rbm = {I \over d'}$.  The
measurement $\cal M$ in the modified protocol should now have an extra
POVM element $I \ot \rho_f^T$ besides those specified in \eq{mm}.

Our result gives insights on when RSP has no advantage over
teleportation.  The oblivious condition causes Bob's quantum state
(given the classical message) to be one obtainable by applying a
quantum operation on $\phi$ (see \eq{rhophim}), which is a necessary
condition for a protocol oblivious to Alice.  
This is in accord with the fact that most RSP protocols are not
oblivious.  We describe an example of an RSP protocol that uses only
forward communication, and is faithful and oblivious to Bob, but works
for a nongeneric ensemble.  Each member in the ensemble is given by
$\phi = {1 \over 2} (I + \cos \theta \cos \eta \; \sigma_x + \cos
\theta \sin \eta \; \sigma_y + \sin \theta \; \sigma_z)$ where
$\sigma_{x,y,z}$ denote the Pauli matrices, each $\eta$ can specify a
member in the ensemble, and $\theta$ denotes a constant in
$[0,\pi/2]$.  In other words, these states lie on a latitude on the
Bloch sphere.  In the RSP protocol, Alice and Bob share $1$ ebit per
qubit prepared.  To transmit $\phi$, Alice performs a trinary
measurement with measurement operators $M_0 = (1-p) \phi^T$, $M_1 =
(1-p) (\sigma_z \phi \, \sigma_z)^T$, and $M_2 = p |1\>\<1|$, where $p
= \sin \theta / (1 + \sin \theta)$ and $M_0 + M_1 + M_2 = I$.  When
the measurement outcome is $m=0,1,2$, Bob obtains $\phi$, $\sigma_z
\phi \, \sigma_z$, and $|1\>\<1|$ respectively, and in the last case,
Alice and Bob perform teleportation.  When $n$ qubits are to be
prepared for large $n$, Alice can apply deterministic data
compression~\cite{dc} with variable message length.  The average
classical communication cost is $n (H(p) + p + 1)$ which is less than
$2 n$ for small $p$ (latitudes close to the equator).  This example
illustrates that a generic ensemble is needed for our result to hold.


We conclude with some open questions.  Extensions to an RSP protocol
that starts with some other entangled state rather than the EPR state,
or one that uses back communication are interesting to consider.

We thank Charles Bennett, David DiVincenzo, Andreas Winter, John
Smolin, and Barbara Terhal for enlightening discussions, and Charles
Bennett for helpful comments on the manuscript.  This research is
supported in part by the NSA and ARDA under the US Army Research
Office, grant DAAG55-98-C-0041.



\end{document}